\documentclass[aip,rsi,reprint,graphicx]{revtex4-1}
\usepackage{CJK}
\usepackage{graphicx}
\usepackage{amsmath}

\begin{document}
\begin{CJK*}{UTF8}{}
\CJKfamily{min}
\title{500\,MHz resonant photodetector for high-quantum-effciency, low-noise homodyne measurement}
\author{Takahiro Serikawa (芹川昂寛) and Akira Furusawa (古澤明)}
\email[]{Author to whom correspondence should be addressed: akiraf@ap.t.u-tokyo.ac.jp}
\affiliation{Department of Applied Physics, School of Engineering, The University of Tokyo, 7-3-1 Hongo, Bunkyo-ku,
Tokyo, 113-8656, Japan.}

\date{\today}

\begin{abstract}
We design and demonstrate a resonant-type differential photodetector for low-noise quantum homodyne measurement at 500\,MHz optical sideband with 17\,MHz of bandwidth. By using a microwave monolithic amplifier and a discrete voltage buffer circuit, a low-noise voltage amplifier is realized and applied to our detector. 12\,dB of signal-to-noise ratio of the shot noise to the electric noise is obtained with 5\,mW of continuous-wave local oscillator. We analyze the frequency response and the noise characteristics of a resonant photodetector, and the theoretical model agrees with the shot noise measurement.
\end{abstract}

\maketitle
\end{CJK*}

\section{introduction}
\begin{figure}
\centering
\includegraphics{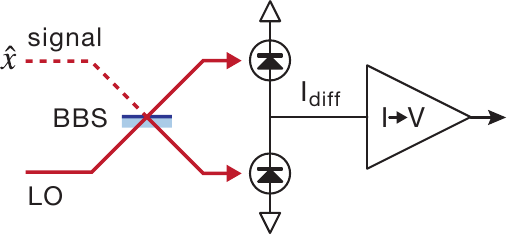}
\caption{\label{fig:hom}Conceptual sketch of balanced homodyne measurement. Signal light interferes with a local oscillator (LO) beam at a balanced beamsplitter (BBS). The differential photo current is obtained by two series photodiodes. An I--V converter circuit extracts a voltage signal.}
\end{figure}

Quantum measurement plays a significant role in the rapidly evolving quantum technologies including quantum sensing, cryptography and computing, since it makes a connection between quantum and classical world. Optical balanced homodyne detection\cite{Yuen:83} is one of the ideal quantum measurements, where nearly unity quantum efficiency and high signal-to-noise ratio (SNR) can be obtained, leading to the applications in gravitational wave detectors\cite{fritschel2014balanced}, quantum-enhanced optical phase tracking\cite{yonezawa2012quantum}, squeezed state detection\cite{breitenbach1997measurement}, quantum state reconstruction\cite{smithey1993complete,lvovsky2009continuous}, and measurement-based quantum operations\cite{PhysRevLett.90.117901,PhysRevLett.80.869}. In homodyne detection, the target light interferes with local oscillator (LO) beam and the quadrature field amplitude $\hat{x}$ is demodulated from the light frequency to an electrical signal (Fig.~\ref{fig:hom}). Here, for the future frequency-division multiplexing of quantum optics\cite{PhysRevLett.112.120505, song2014quantum}, high-frequency detectors are desired to do high-precision quadrature measurements at optical sidebands. Also, they will be used in optical readout of microwave quantum states\cite{bochmann2013nanomechanical}, which is expected to contribute to superconducting quantum computers.

Conventional wideband homodyne detectors for quantum optics have the bandwidth typically below 100\,MHz\cite{1367-2630-13-1-013003,KUMAR20125259}. The bandwidth limit comes from the terminal capacitance of photodiodes, which is particularly a problem for detectors with large-area photodiodes. For quantum homodyne measurement, however, large-area photodiodes are preferred because they can receive the input light beam with nearly unity efficiency, and are tolerant to saturation caused by the high-power local oscillator. The terminal capacitance also deteriorates the SNR in high-frequency region\cite{0256-307X-30-11-114209}, making it difficult to simultaneously achieve high SNR and wide bandwidth. An alternative approach is the resonant architecture\cite{Darcie1988,doi:10.1063/1.2735559,doi:10.1063/1.4966249,doi:10.1063/1.5004418}, where a high transimpedance gain and good SNR is achieved at the resonant frequency. Resonant detectors are often used for high-sensitivity purposes, for example gravitational wave detectors\cite{doi:10.1063/1.2735559}. It is also possibly suitable for high-frequency application for which wideband detectors cannot meet the requirements. Recently Chen {\it et al}. have reported a resonant-type homodyne receiver at 100\,MHz\cite{doi:10.1063/1.4966249}, however, their design is based on a magnetic core inductor and an operational amplifier, which makes it difficult to be applied in microwave frequencies.

We develop a resonant balanced homodyne detector with the center frequency $f_\mathrm{res} = 500\,\mathrm{MHz}$ and 12\,dB of SNR between the shot noise and the electric noise. We make use of a Si-photodiode with 99\% quantum efficiency for 860\,nm light, which makes our detector compatible with conventional quantum optical experiments or quantum measurement of Cs atom D2 line experiments. The key feature of our detector is a resonant structure with an air-core inductor, followed by a combination of a discrete buffer and a monolithic low-noise amplifier (LNA).

In this paper, we briefly summarize the concept of shot noise SNR and give an analysis of wideband photodetectors and resonant detectors. Then we propose our circuit design of a resonant detector and describe the construction method for the high-frequency operation. We check the characteristics of the resonant circuit by directly measuring the impedance spectrum. The high-frequency roll-off of the photodiode is specified in the same condition as the actual use. Finally, the gain spectrum and the shot noise SNR of our detector is characterized by homodyne measurement of a vacuum.

\section{Theory}
\subsection{Signal to noise ratio}
Noise performance of homodyne detection in quantum regime is often evaluated by the SNR of the shot noise to the circuit noise. Shot noise originates from the random arrival of the LO photons, which represents the quantum fluctuation of the quadrature operator of the target optical mode\cite{Yuen:83}. When the signal beam is a vacuum, the differential photon number has a white-noise-like probability statistics, and the spectrum density of the homodyne detection is given as
\begin{align}
 V_\mathrm{shot}(\omega) = 2 q^2\frac{\eta_\mathrm{PD}P_\mathrm{LO}}{h\nu}\, |Z_\mathrm{T}(\omega)|^2,
 \label{eq:shotnoisetr}
\end{align}
where $q$ is elementary charge, $\eta_\mathrm{PD}$ is the quantum efficiency of the photodiode, $P_\mathrm{LO}$ is photo-power of the LO, $h$ is Plank's constant, $\nu$ is the frequency of the LO, and $Z_\mathrm{T}(\omega)$ is the transimpedance gain of the detector. This is compared with the variance of the electric noise $V_\mathrm{elec}(\omega)$, giving the signal-to-noise ratio (SNR) of shot noise:
\begin{align}
\mathcal{R}(\omega) = \frac{V_\mathrm{shot}(\omega)}{V_\mathrm{elec}(\omega)}.
\label{eq:sn}
\end{align}
Note that with the same electric noise level, the shot noise SNR gets better as the LO power increase, yet the saturation of the photodiodes gives rise to a limit of the dynamic range.

Quantum quadrature is derived from the voltage signal of homodyne detection normalized by $V_\mathrm{shot}$. In this sense, we can interpret the shot noise as a vacuum fluctuation\cite{Yuen:83}. There are two factors which restrict the measurement precision: the electric noise and the finite quantum efficiency of photodiodes. It is known that these two are equally treated as optical losses\cite{PhysRevA.75.035802}, since the vacuum fluctuation and the Gaussian electric noise have the same probability distribution. The frequency-dependent effective optical efficiency is written as
\begin{align}
 \eta_\mathrm{eq}(\omega) = \frac{\mathcal{R}(\omega)}{1 + \mathcal{R}(\omega)}.
 \label{eq:eqeta}
\end{align}
Under the existence of the electronic noise and finite quantum efficiency of the photodiode $\eta_\mathrm{PD}$, the output signal of homodyne detection is interpreted as a quadrature measured with total efficiency of $\eta_\mathrm{PD}\eta_\mathrm{eq}$. The frequency spectrum of the measured quadrature reads
\begin{align}
\begin{aligned}
\hat{x}_\mathrm{measured}(\omega) = &\sqrt{\eta_\mathrm{PD}\eta_\mathrm{eq}(\omega)}\,\hat{x}_\mathrm{state}(\omega)\\ &\hspace{2em} + \sqrt{1 - \eta_\mathrm{PD}\eta_\mathrm{eq}(\omega)}\,\hat{x}_\mathrm{vac}(\omega),
\end{aligned}
\end{align}
where $\hat{x}_\mathrm{state}$ is the quadrature operator of the target field and $\hat{x}_\mathrm{vac}$ is a Gaussian noise term represented as vacuum fluctuation. Therefore, to extract the full information of the quantum state $\hat{x}_\mathrm{state}$, the electric noise must be much smaller than the shot noise level and $\eta_\mathrm{PD}$ must be close to one.

\subsection{Wideband detectors}
\begin{figure}
\centering
\includegraphics{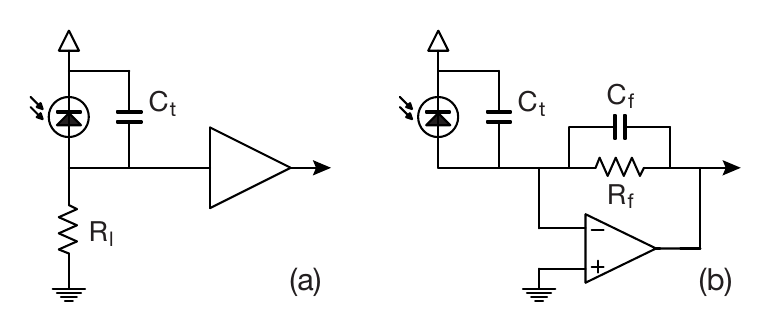}
\caption{\label{fig:statia} Circuit models of (a) a load-resister detector and (b) a feedback TIA.}
\end{figure}
Before describing resonant photodetectors, we would like to glance at standard photodetectors, whose gain spectrum spans from 0\,Hz. We show that it is hard to realize a high-SNR homodyne detector at 500\,MHz band with the wideband architecture employing state-of-the-art operational amplifiers.

Figure \ref{fig:statia} shows the schematics of these amplifiers, where photodiodes are modeled as a photocurrent source paralleled with a terminal capacitance $C_\mathrm{t}$ which cause a limit of bandwidth and SNR. Note that the input capacitance of the amplifier is included in $C_\mathrm{t}$.

In a load-resistor detector, the photocurrent $I_\mathrm{photo}$ generates signal voltage $V_\mathrm{sig}$ when it goes through $C_\mathrm{t}$ and load resistance $R_\mathrm{l}$. The transimpedance gain is given by
\begin{align}
 Z_\mathrm{load \mathchar`- R}(\omega) = \frac{R_\mathrm{l}}{1 + i\omega C_\mathrm{t} R_\mathrm{l}}.
 \label{eq:loadz}
\end{align}
Assuming that we can ignore the dark current of the photodiode and the voltage noise of the buffer, only the thermal noise of the transimpedance contributes to the electric noise spectrum:
\begin{align}
\begin{aligned}
 V_\mathrm{load \mathchar`- R}(\omega) = 4k_\mathrm{B} T\, \mathrm{Re}\bigl[Z_\mathrm{load \mathchar`- R}\bigr]= \frac{4k_\mathrm{B} T R_\mathrm{l}}{1 + C_\mathrm{t}^2R_\mathrm{l}^2\omega^2},
 \label{eq:loadrnoise}
\end{aligned}
\end{align}
where $k_\mathrm{B}$ is Boltzman's constant and $T$ is the temperature of the load resistor. From Eqs.~(\ref{eq:sn}), (\ref{eq:loadz}), (\ref{eq:loadrnoise}), the shot noise SNR is given by
\begin{align}
\mathcal{R}_\mathrm{load \mathchar`- R}(\omega) = \frac{q^2\eta_\mathrm{PD}P_\mathrm{LO}}{2k_\mathrm{B}Th\nu}R_\mathrm{l}.
\end{align}
Here a trade-off appears between the gain bandwidth and the SNR. Larger $R_\mathrm{l}$ improves the SNR, while narrowing the gain bandwidth which is characterized by the pole frequency $f_\mathrm{c} = 1 / (2\pi R_\mathrm{l}C_\mathrm{t})$. Note that the SNR spectrum of this type detectors does not depends on frequency. However, to extract the small signal in high-frequency region, unfeasible requirements are placed on the noise-level of the following amplifier.

To expand the bandwidth of load-resistor detectors, feedback architecture is applied as in Fig.~\ref{fig:statia}(b). The operational amplifier with a voltage gain of $A(\omega)$ reduces the input impedance as
\begin{align}
 Z_\mathrm{in}(\omega) = \frac{Z_\mathrm{f}(\omega)}{1-A(\omega)},
\end{align}
where $Z_\mathrm{in}(\omega)$ is the imput impedance and $Z_\mathrm{f} = R_\mathrm{f}/ (1 + i\omega R_\mathrm{f}C_\mathrm{f})$ is the feedback impedance of the TIA. We avoid the complicated discussion about the loop-stability and response characteristics of feedback TIAs. From the detailed calculations\cite{masalov2017noise}, the gain-bandwidth of a feedback TIA is arbitrarily expanded as long as the operational amplifier is fast enough, leading to wideband transimpedance gain spectrum. However, the SNR of feedback TIAs gets worse at higher frequency, since the noise spectrum suffer from the voltage noise of the operational amplifiers, which is amplified in high-frequency region by the feedback structure. The input voltage noise spectrum density of the amplifier $e_n$ contributes to the noise spectrum as:
\begin{align}
 V_\mathrm{TIA}(\omega) =  \frac{4k_\mathrm{B}T}{R_\mathrm{f}}|Z_\mathrm{TIA}(\omega)|^2 + \bigl|\omega C_\mathrm{t}\,Z_\mathrm{TIA}(\omega) + 1 \bigr|^2 e_n^2,
\end{align}
where we assume that the input current noise of the amplifier is negligible (it can be canceled by adding an external low-noise JFET buffer \cite{0256-307X-30-11-114209}). The second term becomes dominant in high-frequency domain. The frequency dependance of this term can be interpreted as follows; $e_n$ is once converted to a current noise term which leaks through the terminal capacitance, and then it appears in the output voltage derived by the TIA. Assuming $\bigl|\omega C_\mathrm{t}Z_\mathrm{TIA}(\omega)\bigr| \gg 1$, the SNR of a feedback TIA is calculated as,
\begin{align}
 \mathcal{R}_\mathrm{TIA}(\omega) = \frac{q^2\eta_\mathrm{PD}P_\mathrm{LO}}{2k_\mathrm{B}Th\nu}\left[\frac{1}{R_\mathrm{f}} + \frac{\omega^2C_\mathrm{t}^2e_n^2}{4k_\mathrm{B}T}\right]^{-1}.
 \label{eq:sntia}
\end{align}
Thus $e_n$ becomes a dominant noise source in high-frequency region and limits the SNR.

To examine the possibility of realizing a low-noise wideband photodetector whose gain bandwidth spans up to 500MHz, we calculate the SNRs of two design of photodetectors. Here we discuss the balanced-homodyne detection where two photodiodes are connected in series to subtract the two photocurrents, whose terminal capacitance is assumed to be 2.0\,pF respectively with the LO power of 5\,mW at 860\,nm wavelength. We assume the use of wideband operational amplifier LMH6629 (Texas Instruments). From the datasheet, it has an exceptionally low voltage noise level $e_n = 0.69\,\mathrm{nV}/\sqrt{\mathrm{Hz}}$, and input capacitance of 5.7\,pF. Then the shot noise SNR from Eq.~(\ref{eq:sntia}) falls down to 2.2\,dB at 500\,MHz, almost regardless of the feedback resistance.

\subsection{Resonant detectors}

\begin{figure}
\centering
\includegraphics{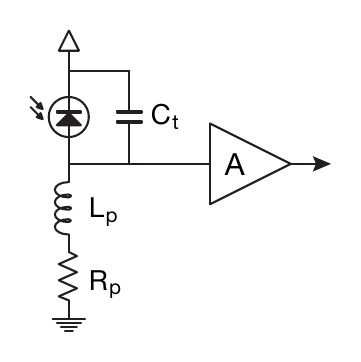}
\caption{\label{fig:restia} Schematic of a resonant-type photodetector.}
\end{figure}

The conceptual schematic of a resonant photodetector is shown in Fig.~\ref{fig:restia}. A parallel inductance $L_\mathrm{p}$ forms a L--C resonant circuit together with the terminal capacitance $C_\mathrm{t}$. The transimpedance $Z_\mathrm{T}$ is calculated as
\begin{align}
 Z_\mathrm{T}(\omega) = \frac{R_\mathrm{p}^2+\omega^2L_\mathrm{p}^2}{R_\mathrm{p} + i\omega(R_\mathrm{p}^2C_\mathrm{t}+\omega^2L_\mathrm{p}^2C_\mathrm{t}-L_\mathrm{p})},
 \label{eq:transz_res}
\end{align}
where a series resistance $R_\mathrm{p}$ of the inductor is also considered. At the resonance frequency
\begin{align}
 \omega_\mathrm{res} = \sqrt{\frac{1}{L_\mathrm{p}C_\mathrm{t}} - \left(\frac{R_\mathrm{p}}{L_\mathrm{p}}\right)^2},
 \label{eq:omegares}
\end{align}
the peak impedance of
\begin{align}
 R_\mathrm{res} = \frac{L_\mathrm{p}}{R_\mathrm{p}C_\mathrm{t}}
 \label{eq:rres}
\end{align}
is realized. The output voltage is obtained as the product of a photocurrent $I(\omega)$, the transimpedance $Z_\mathrm{T}(\omega)$, and the amplifier's gain $A(\omega)$. We look at the input voltage signal $I(\omega)Z_\mathrm{T}(\omega)$ and omit $A(\omega)$ to simply discuss the SNR of the detector.

Introducing the frequency deviation from the resonance as $\omega_\mathrm{d} = \omega - \omega_\mathrm{res}$, Eq.~(\ref{eq:transz_res}) is expressed as
\begin{align}
 Z_\mathrm{T}(\omega_\mathrm{res}+\omega_\mathrm{d}) = \frac{R_\mathrm{p}^2+(\omega_\mathrm{res}+\omega_\mathrm{d})^2L_\mathrm{p}^2}{R_\mathrm{p} + iL_\mathrm{p}^2C_\mathrm{t}(\omega_\mathrm{res}+\omega_\mathrm{d})(2\omega_\mathrm{res}+\omega_\mathrm{d})\omega_\mathrm{d}}.
\end{align}
When we only look at a narrow frequency range around the resonance, we can assume $\omega_\mathrm{d} \ll \omega_\mathrm{res}$, leading to
\begin{align}
 Z_\mathrm{T}(\omega_\mathrm{res}+\omega_\mathrm{d}) \sim \frac{R_\mathrm{res}}{1 + 2i\left(R_\mathrm{res} -R_\mathrm{p}\right) C_\mathrm{t}\omega_\mathrm{d}}.
 \label{eq:transimpedance_simple}
\end{align}
This is a first-order band-pass response around $\omega_\mathrm{res}$, characterized by the cutoff frequency (i.e. the bandwidth) of
\begin{align}
 f_\mathrm{c} = \frac{1}{2\pi}\frac{1}{2(R_\mathrm{res} - R_\mathrm{p})C_\mathrm{t}}.
 \label{eq:resfc}
\end{align}
Usually $R_\mathrm{p}$ is much smaller than $R_\mathrm{res}$ and can be ignored, leading to $f_c \sim 1/4\pi R_\mathrm{res}C_\mathrm{t}$. Thus there is a trade-off between bandwidth and SNR just like the load-register detector. Since the time-resolution of the detector becomes critical to measure the quantum state of wavepackets with a continuous-wave LO\cite{}, it is preferred to reduce the terminal capacitance to obtain higher bandwidth while keeping the SNR.

The input voltage noise spectrum of a resonant detector is given by a sum of the thermal noise of the transimpedance $Z_\mathrm{T}$ and the voltage noise of the amplifier $e_n$, reading
\begin{align}
 V(\omega_\mathrm{res} + \omega_\mathrm{d}) &\sim  \frac{4k_\mathrm{B} TR_\mathrm{res}}{1 + 4R_\mathrm{res}^2C_\mathrm{t}^2\omega_\mathrm{d}^2} + e_n^2,
 \label{eq:vn_res}
\end{align}
where we assume $R_\mathrm{p} \ll R_\mathrm{res}$.
The effective (noise-free) bandwidth of the resonant detector is limited by the voltage noise of the amplifier, which becomes dominant in the noise level in high frequency region. The thermal noise does not contribute to the frequency-dependence of the SNR by itself, like as load-resistor detectors. The shot noise SNR is written as
\begin{align}
\begin{aligned}
 \mathcal{R}_\mathrm{res}(\omega_\mathrm{res} + \omega_\mathrm{d})
  \sim \frac{2q^2\eta_\mathrm{PD}P_\mathrm{LO}/h\nu}{4k_\mathrm{B}T/R_\mathrm{res} + e_n^2/R_\mathrm{res}^2 + 4C_\mathrm{t}^2e_n^2\omega_\mathrm{d}^2}.
 \label{eq:ressn}
\end{aligned}
\end{align}
The SNR at the resonance frequency gets better when we choose larger $R_\mathrm{res}$ (i.e. smaller $R_\mathrm{p}$). At larger $\omega_\mathrm{d}$, however, the $\omega_\mathrm{d}^2$-proportional term becomes dominant and the SNR is almost independent of $R_\mathrm{res}$. The off-resonance SNR is governed by $C_\mathrm{t}$ and $e_n$, which should be small to obtain a broad noise-free bandwidth.

\section{Design}
\subsection{Design concepts}

First, we summarize the conditions of the detector design. We target the center frequency of 500\,MHz. We choose a specially made Si-PIN photodiode S5971SPL (Hamamatsu Photonics), which has 99\% quantum efficiency at 860\,nm and 2.0\,pF of terminal capacitance. S5971SPL equips a circle photosensitive aperture with a diameter of 0.8\,mm, and has a particularly thick intrinsic layer, assuring the high quantum efficiency and the low terminal capacitance.

\begin{figure*}[htb]
\centering
\includegraphics{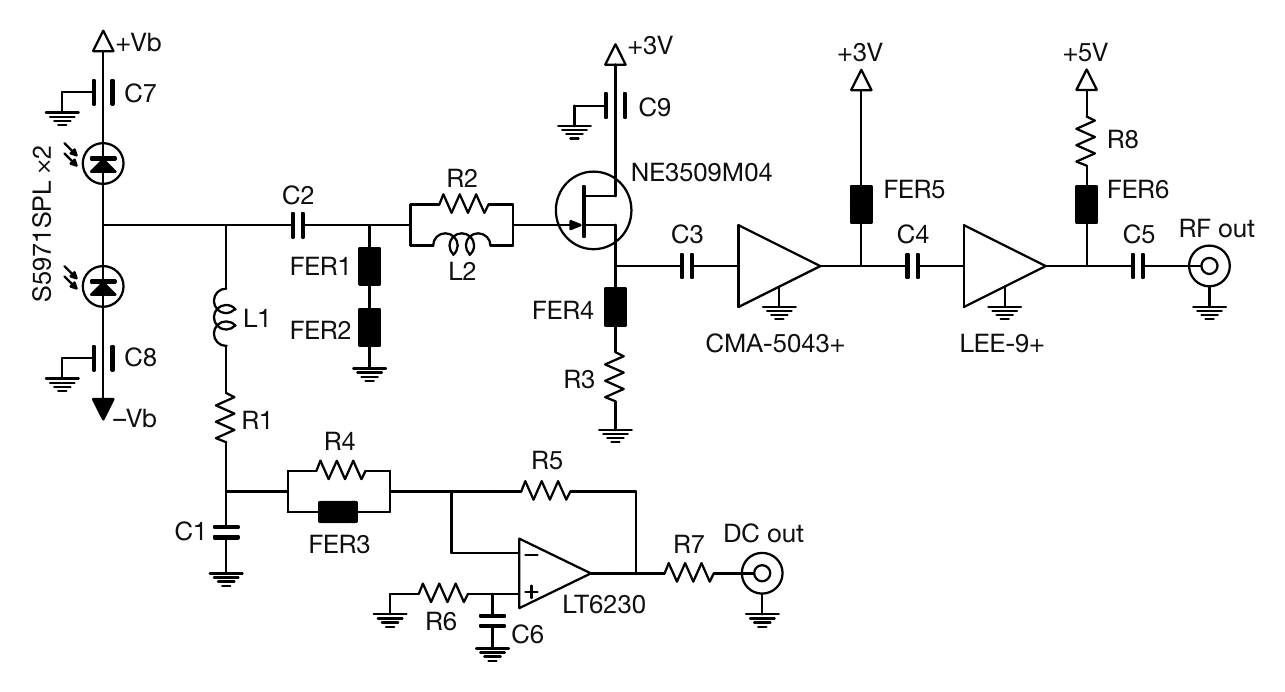}
\caption{\label{fig:cir}Circuit schematic of our detector. FER: ferrite beads. Minor components, such as bypass capacitors, power-line noise filters, or voltage regulators are omitted.}
\end{figure*}

\begin{table}[htb]
  \caption{\label{table:partlist}Part list of the circuit schematic Fig.~(\ref{fig:cir})}
  \begin{tabular}{l|r}
  \hline\hline
   C1&\ 200p (GQM1875C2E101 $\times2$)\\\hline
   C2, C3, C4, C5& 100p (GQM1875C2E101)\\\hline
   C6& 100n\\\hline
   C7, C8& 100n (NFM31KC104)\\\hline
   C9& 10u (NFM21PS106)\\\hline
   L1& 18n (air-core)\\\hline
   L2& 10n (LQW18AN10)\\\hline
   R1& 3.9\\\hline
   R2& 150\\\hline
   R3& 47\\\hline
   R4& 100\\\hline
   R5, R6& 1k\\\hline
   R7& 51\\\hline
   R8& 22\\\hline
   FER1, FER2& BLM18HE152SN1\\\hline
   FER3& BLM18RK102SN1\\\hline
   FER4, FER5, FER6& BLM18HE152SN1\\\hline
  \hline
  \end{tabular}
\end{table}

Figure \ref{fig:cir} shows the circuit schematic of our detector and the part list is presented in Table \ref{table:partlist}. Two series S5971SPL photodiodes receive the two balanced LO beams, transferring the differential photocurrent to the amplifier stage. These photodiodes are anti-biased at $V_\mathrm{b} = 180\,\mathrm{V}$ so that the transit-time is minimized (discussed in Sec.~\ref{sec:pdcutoff}).

The transimpedance spectrum $Z_\mathrm{T}(\omega)$ is determined by the choice of the parallel inductance $L_\mathrm{p}$ and the series resistance $R_\mathrm{p}$. The terminal capacitance $C_\mathrm{t}$ is estimated at 4.6\,pF, considering the capacitance of two photodiodes (4.0\,pF) and the input capacitance of the amplifier (0.6\,pF, explained later).We set $L_\mathrm{p} (L1)=22\,\mathrm{nH}$ and $R_\mathrm{p} (R1)=3.9\,\Omega$, expecting $f_\mathrm{res} = 5.0\times 10^{2}\,\mathrm{MHz}$, $R_\mathrm{res} = 1.2\times 10^{3}\,\Omega$, from Eqs.~(\ref{eq:omegares}) and (\ref{eq:rres}) respectively.

The choice of a voltage amplifier is critical to obtain a good SNR and large bandwidth. The following requirements would be imposed on the amplifier:
\begin{enumerate}
\item The input impedance is high, not to reduce $R_\mathrm{res}$ and increase $C_\mathrm{t}$.
\item The input noise is extremely small such that the small off-resonant voltage signal is extracted with high SNR, since the noise-free bandwidth is determined by the off-resonant SNR.
\item The voltage gain is high enough around the resonant frequency so that the output signal is easily treated.
\end{enumerate}
Here we make use of CMA-5043+ (Mini-Circuits) for the first-stage amplifier. It is a monolithic microwave integrated circuit with the gain of 22\,dB and the noise figure (NF) of 0.75\,dB (equivalently, the input voltage noise density of $0.20\,\mathrm{nV}/\sqrt{\mathrm{Hz}}$). A problem of such kind of microwave amplifiers is the low input impedance expected to be matched to $50\,\Omega$ transmission lines, since it limits $R_\mathrm{res}$. CMA-5043+ is internally matched to $50\,\Omega$ over the working frequency band and difficult to be directly applied to resonant detectors. We add a voltage follower to drive the input of CMA-5043+. It is a source follower composed of NE3509M04 (CEL), which is a high-speed, very low-noise ($\mathrm{NF} = 0.4\,\mathrm{dB}$, noise density of $0.28\,\mathrm{nV}/\sqrt{\mathrm{Hz}}$) hetero-junction field effect transistor (HJFET). NE3509M04 replaces the input impedance of CMA-5043+ by its gate-to-drain capacitance $C_\mathrm{gd} = 0.6\,\mathrm{pF}$, which is to be included in $C_\mathrm{t}$. JFETs (including HJFETs) are rather preferred than bipolar transistors here because the shot noise of base current of bipolar transistors is considerably large for this purpose. A gate resistance R2 is inserted to avoid an oscillation of the source follower at several GHz region and the parallel inductor L2 reduces the thermal noise of R2 at 500\,MHz. To set the bias condition of NE3509M04, the gate voltage is fixed at 0\,V by connecting it to ground via the ferrite beads FER1, 2. The bias current of the follower circuit is 10\,mA, which is settled by the source resistance R3. In total, the first stage amplifier composed of NE3509M04 and CMA-5043+ equips high input-impedance ($Z_\mathrm{in}(\omega) = 1/i\omega C_\mathrm{gd}$), wide bandwidth (4\,GHz), low input-noise ($0.35\,\mathrm{nV}/\sqrt{\mathrm{Hz}}$), and 28\,dB of voltage gain. The second amplifier LEE-9+ have a flat 9\,dB gain over several GHz band and provides good output matching to $50\,\Omega$ transmission lines.

The low-frequency photocurrent is subtracted by a transimpedance amplifier made of a high-accuracy operational amplifier LT6230 (Linear Technology). Not to affect the resonance characteristics around 500\,MHz, the low-frequency TIA is isolated by resistor R4. Bypass capacitor C1 shunts the high-frequency component to the ground.

\subsection{Implementation}

\begin{figure*}[htb]
\centering
\includegraphics{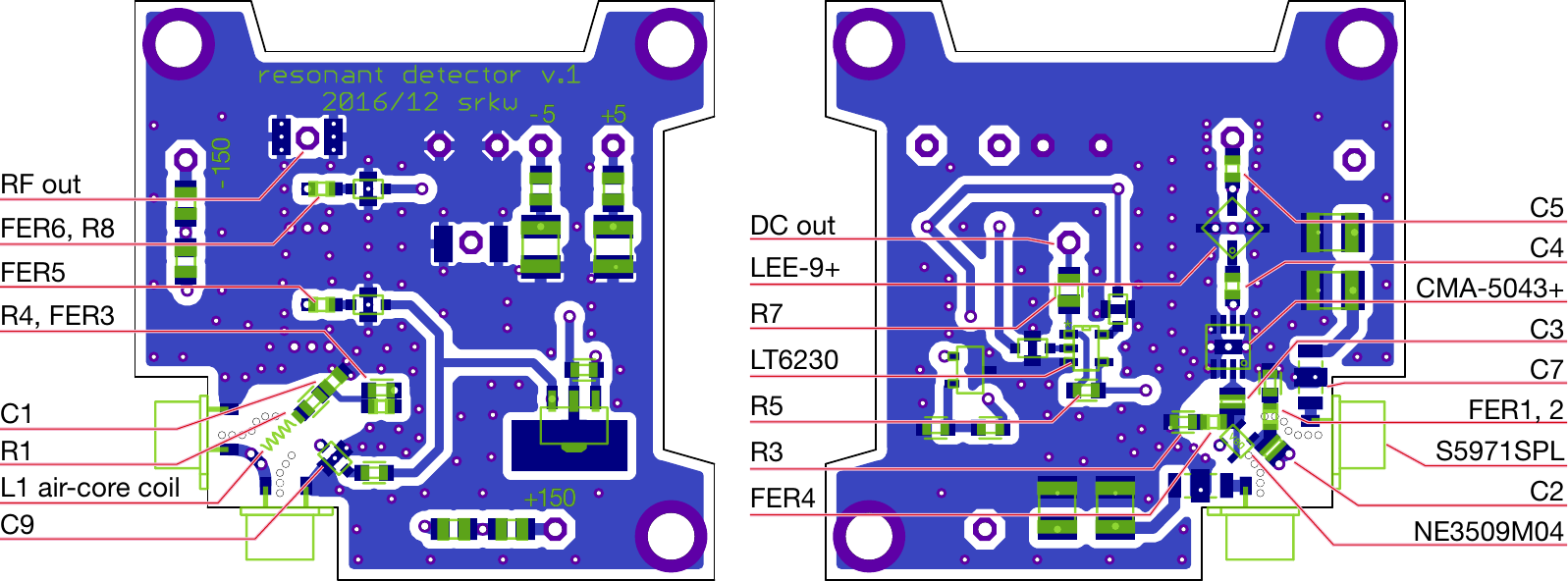}
\caption{\label{fig:pcb}PCB board layout. Left: top layer. Right: bottom layer.}
\end{figure*}

We carefully designed the printed circuit board (PCB) layout to reduce unwanted stray capacitance or inductance (Fig.~\ref{fig:pcb}). The PCB is a two-layer, 1.6\,mm thick composite board (FR-4). S5971SPL photodiodes are mounted at the edge of the PCB so that the junction node of the two photodiodes is kept small and well isolated from the ground patterns, and also the optical path of the homodyne detection is easily routed. Note that L2 and R2 are not on the layout because they are added ad hoc.

\begin{figure}[htb]
\centering
\includegraphics{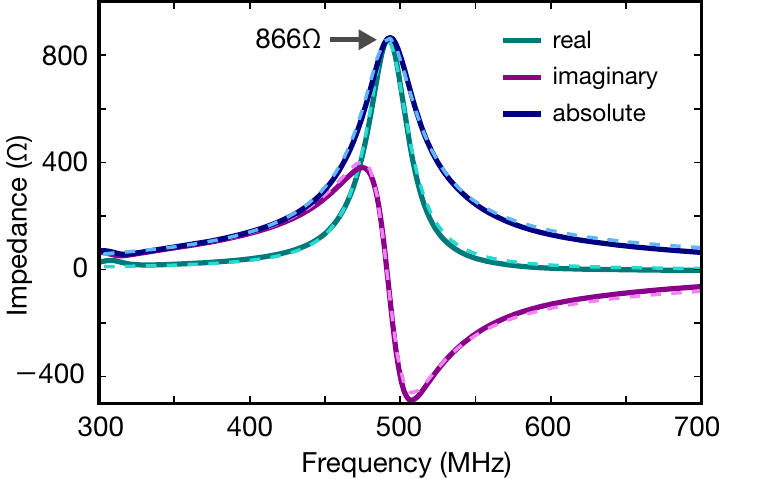}
\caption{\label{fig:impedance} Transimpedance spectrum of our resonant detector. Dashed lines indicates the fitting curve from Eq.~(\ref{eq:transimpedance_simple})}
\end{figure}

The resonant inductor L1 is a homemade air-core coil made of a $\phi 0.3\,\mathrm{mm}$ tin-coated copper wire, which is winded five times with a diameter of 2.0\,mm. The inductance can be tuned within few nH range by stretching the length. During the adjustment, the resonance is directly monitored by a network analyzer (Keysight E5061B). Here a $50\,\Omega$ coaxial cable is temporary soldered at the junction node of the two photodiodes, and the reflection coefficient ($\mathrm{S}_{11}$) is observed. Figure~\ref{fig:impedance} shows the frequency spectrum of $Z_\mathrm{T}$ calculated from the S-parameter $\mathrm{S}_{11}$, obtaining $f_\mathrm{res}=491\,\mathrm{MHz}$, $R_\mathrm{res} = 866\,\Omega$ and $f_\mathrm{c} = 17\,\mathrm{MHz}$. The fitting with Eq.~(\ref{eq:transimpedance_simple}) predicts $C_\mathrm{t} = 5.5\,\mathrm{pF}$, and $L_\mathrm{p} = 19\,\mathrm{nH}$. Thus we estimate an excess capacitance at 0.9\,pF, which we suppose to be the parasitic capacitance of the PCB pattern.

\begin{figure}[htb]
\centering
\includegraphics{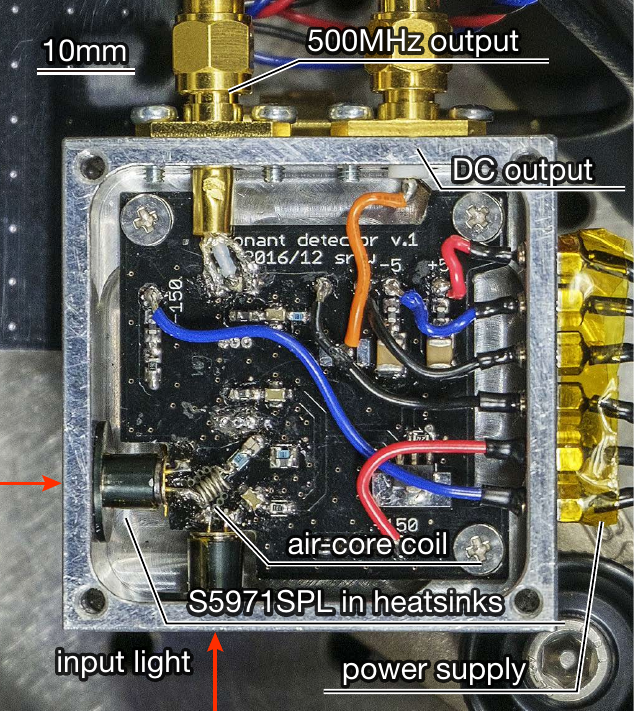}
\caption{\label{fig:pic}Picture of the detector, placed on a optical table.}
\end{figure}

In order to protect the detector circuit from electromagnetic noise, we designed a  machined aluminum box to isolate the circuit (Fig.~\ref{fig:pic}). This box defines a steady ground reference and makes an electric contact with the ground pattern of the PCB. For the power line inputs, feed-through capacitors (Kondo Electric FSA30250F102P) are inserted to reject the electromagnetic noise and the switching noise of the power supply. This box also works as a heatsink of the photodiodes, where cap-type heatsinks (Aavid Thermalloy 322400B00000G) makes a thermal contact between the photodiodes and the box. The low thermal resistance suppresses the reverse leakage current of photodiodes, which increases exponentially with the temperature\cite{Fraser86}.

\section{Benchmark}
\subsection{Cutoff and saturation of S5971SPL}
\label{sec:pdcutoff}
\begin{figure}[htb]
\centering
\includegraphics{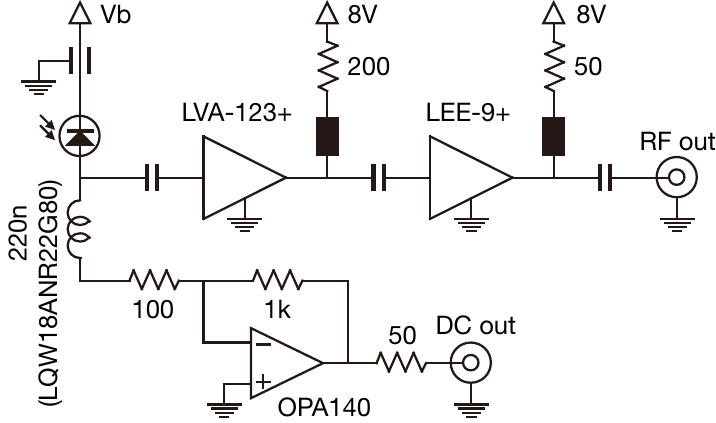}
\caption{\label{fig:cirwide}Circuit schematic of the wideband photodetector used for the specification of S5971SPL.}
\end{figure}

We first specify the high-frequency characteristics of the photodiodes. Since S5971SPL equips very thick intrinsic region for the high quantum efficiency, the large transit-time of the photocarriers\cite{doi:10.1063/1.1713426} cause a roll-off at several hundred MHz. Especially, the space-charge effect\cite{dentan1990numerical,williams1994effects,harari1996modeling} lowers the transit-time cutoff when a high-power LO is used to improve the SNR. Thus higher bias voltage is preferred to reduce the transit-time; this is why we apply as high as 180\,V to S5971SPL.

The high-frequency response of S5971SPL is experimentally identified by measuring the shot noise spectrum. We use a wideband photodetector circuit shown in Fig.~\ref{fig:cirwide}. This is a load-resister photodetector, where the internally-matched microwave amplifier LVA-123+ (Mini-Circuits) provides $50\,\Omega$ transimpedance and 23\,dB of flat-response amplification over several GHz. The cut-off frequency formed by the terminal capacitance of S5971SPL and the $50\,\Omega$ load resistance is calculated to be 1.5\,GHz from Eq.~(\ref{eq:loadz}), which is negligible in comparison with the transit-time cutoff.

\begin{figure}[htb]
\centering
\includegraphics{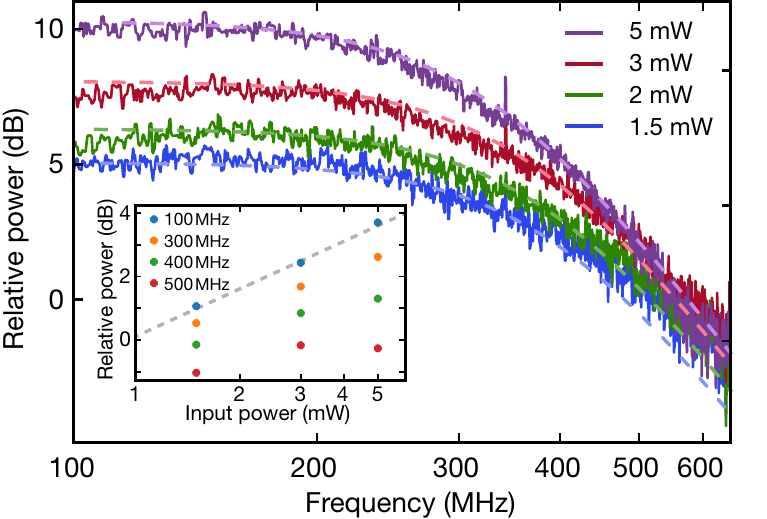}
\caption{\label{fig:s5971sat} Shotnoise responce spectrum of S5971SPL, measured by using the circuit in Fig.~\ref{fig:cirwide}.  The x-axis is logarithmic scale. The input light is continuous-wave single-mode laser at 860\,nm. The electric noise of the detector circuit is subtracted from each traces. Theoretical curves of the shotnoise spectrum (shown as dashed lines) are modeled by Eq.~(\ref{eq:satmodel}). Sub-plot: input power dependence of the detected shotnoise power. The x-axis is logarithmic scale. The gray dashed line shows the predicted shotnoise power, which linearly increases with the input light power. At high input power, the effect of transit-time cutoff gets larger due to the saturation of the photodiodes.}
\end{figure}

Figure \ref{fig:s5971sat} shows the measured shot noise spectra of S5971SPL with several LO power under the bias voltage of 180\,V. The LO light makes 0.5\,mm diameter circular spots on the photodiodes. Since shot noise has a white spectrum in frequency-domain, these traces can be considered as the gain spectrum of S5971SPL. At 500\,MHz, the detected power barely gets larger as the LO intensity increases above 3\,mW.

We fit the roll-off of the photodiode with a 2nd-order low-pass response
\begin{align}
 \mathcal{F}(\omega) = \frac{1}{1 + 2i\kappa\omega/\omega_\mathrm{pd}(P) - \bigl[\omega/\omega_\mathrm{pd}(P)\bigr]^2},
 \label{eq:satmodel}
\end{align}
where $\omega_\mathrm{pd}(P)$ is the cutoff frequency depending on the input power $P$ and $\kappa$ is the damping factor. The fitting results in the estimated parameter of $\kappa=0.7$ and the input-power-dependent cutoff frequency $f_\mathrm{pd} = \omega_\mathrm{pd}/2\pi$ as a linear function of the light power as
\begin{align}
 f_\mathrm{pd}(P) = 420\,\mathrm{MHz} - P\times19\,\mathrm{MHz/mW}.
\end{align}
 We emphasize that this is a phenomenological model without referring to the complex physical mechanisms like the carrier transport dynamics in Si.

\subsection{Measurement of shot noise}
The gain and SNR spectrum of our detector is obtained by measuring the shot noise. The light source is a continuous-wave Ti:Sapphire laser (Coherent MBR-110) operating at 860\,nm. The LO light is split by a 50:50 beamsplitter and incidents into the photodiodes as the 0.44\,mm diameter spots. We tune the split ratio of the beamsplitter to cancel the DC differential photocurrent. The path-length of each beam is balanced by canceling an amplitude modulation at 500\,MHz in the LO which is temporary applied during the alignment process, obtaining 32\,dB of common-mode rejection. The signal spectrum is acquired by a spectrum analyzer (Agilent E4402B) with the resolution bandwidth of 100\,kHz and averaged over 1000 times. We slightly adjusted the air-core coil L1 by stretching it on-site so that the shot noise spectrum has the peak at 500\,MHz. The optical power of the LO is set at 5.0\,mW (thus, 2.5\,mW for each photodiode).

\begin{figure}[htb]
\centering
\includegraphics{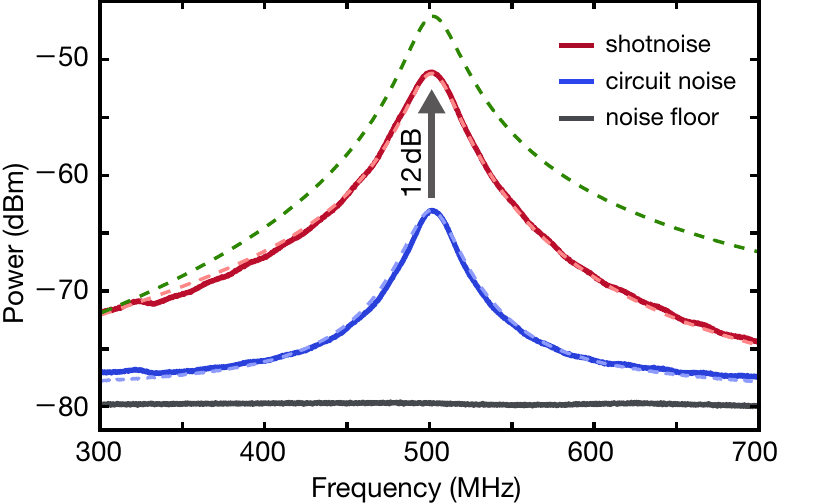}
\caption{\label{fig:shotnoise} Shotnoise and circuit noise spectrum of our detector. Red trace: shot noise of 5.0\,mW of LO; blue trace: circuit noise; gray trace: noise floor of E4402B; green dashed line: theoretical prediction of shot noise spectrum without the roll-off of S5971SPL.}
\end{figure}

Figure \ref{fig:shotnoise} shows the power spectrum of the shot noise and the circuit noise. We obtained 12\,dB of shot noise SNR at 500\,MHz. A theoretical curve of shot noise spectrum is calculated from Eqs.~(\ref{eq:shotnoisetr}), (\ref{eq:transimpedance_simple}), and (\ref{eq:satmodel}), where the adjusted $L_\mathrm{p}$ is estimated at $18\,\mathrm{nH}$. We also show a theoretical prediction of the shot noise where the transit-time characteristics of S5971SPL is ignored. It indicates there is 6\,dB of suppression in the signal gain and SNR at 500\,MHz due to the roll-off of S5971SPL. The circuit noise spectrum has a good agreement with Eq.~(\ref{eq:vn_res}). In the theoretical curve of the circuit noise, we included the noise floor of the spectrum analyzer also shown in Fig.~\ref{fig:shotnoise}.

\begin{figure}[htb]
\centering
\includegraphics{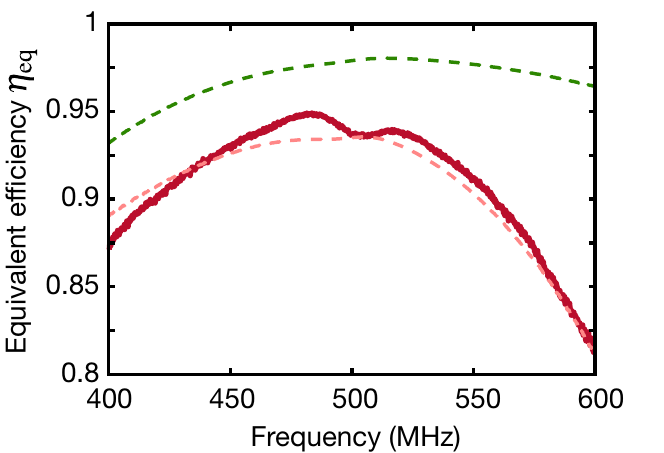}
\caption{\label{fig:eqeff} Equivalent optical efficiency spectrum of the homodyne measurement with the LO power of 5.0\,mW. Red dashed trace: theoretical curve, and green dashed trace: theoretical prediction without the roll-off of S5971SPL.}
\end{figure}

The equivalent optical efficiency is calculated from Eq.~(\ref{eq:eqeta}) and shown in Fig.~\ref{fig:eqeff}. A low-noise homodyne measurement with the equivalent efficiency above 0.9 is realized from 420\,MHz to 560\,MHz, thanks to the low input-noise of the first stage amplifier. Although the gain-bandwidth is 17\,MHz as expected, we have a good SNR in broader bandwidth, which enables a precise quantum measurement of broadband wavepackets around 500\,MHz sideband. The roll-off of S5971SPL reduces the efficiency from 0.98 to 0.93 at 500\,MHz, and a photodiode with a thinner intrinsic layer would realize better total efficiency at 500\,MHz, even though such device will have smaller quantum efficiency.

Currently, the working frequency of our detector is bounded by the cutoff of S5971SPL; and we do not have other indications that limit the resonant frequency. Using faster photodiodes, for example InGaAs photodiodes for longer wavelength light, potentially higher-frequency (several GHz) resonant detectors would be realized with the proposed circuit architecture.

\section{Conclusion}
We have made a comparison of wideband photodetectors and resonant detectors. Resonant detectors have larger SNR at the resonance frequency, while requiring high-frequency, low-noise, high-impedance amplifiers. We have designed a resonant photodetector with the center frequency of 500\,MHz and the bandwidth of 17\,MHz, employing a microwave monolithic integrated circuit and a discrete voltage buffer. Our detector is specified in detail referring to the roll-off characteristics of the photodiodes, the resonance of the load impedance, and the shot noise spectrum. 12\,dB of shot noise SNR is obtained at 500\,MHz with 5.0\,mW of LO power, showing the possibility of high-quantum-efficiency quantum measurements at 500\,MHz optical sideband.

\section{Acknowledgements}
This work was partly supported by CREST (JPMJCR15N5) of JST, JSPS KAKENHI, and APSA of Japan.

%

\end{document}